\documentclass[nonacm=true,sigconf]{acmart}

\usepackage{amsfonts}
\usepackage{mathcomp}
\usepackage{multirow}
\usepackage{pifont}

\usepackage{caption}
\usepackage{subcaption}
\usepackage{tabularx}

\AtBeginDocument{%
  \providecommand\BibTeX{{%
    \normalfont B\kern-0.5em{\scshape i\kern-0.25em b}\kern-0.8em\TeX}}}

\setcopyright{acmcopyright}
\copyrightyear{2018}
\acmYear{2018}
\acmDOI{10.1145/1122445.1122456}

\acmConference[ICPC 2022]{The 30th  International Conference on Program Comprehension}{May 21–22, 2022}{Pittsburgh, PA, USA}
\acmBooktitle{ICPC '22: Proceedings of the 30th  International Conference on Program Comprehension, May 21–22, 2022, Pittsburgh, PA, USA}
\acmISBN{978-1-4503-XXXX-X/18/06}

\theoremstyle{acmplain}

\begin{document}

%\title{An Empirical Study of Comprehension of Feature Models Using Eye-trackers}

%\title{Analysis of Comprehension of Feature Models With Eye-Trackers: An Empirical Study}

\title{An Empirical Eye-Tracking Study of Feature Model Comprehension}

%\title{Analyzing Feature Model Comprehension With Eye-Trackers: \\ An Empirical Study}


\author{Elmira Rezaei Sepasi}
\email{elmira.rezaei-sepasi.1@ens.etsmtl.ca}
\author{Kambiz Nezami Balouchi}
\email{kambiz.nezami-balouchi.1@ens.etsmtl.ca}
\affiliation{%
  \institution{\'Ecole de technologie sup\'erieure, Universit\'e du Qu\'ebec}
  \streetaddress{1100 rue Notre-Dame Ouest}
  \city{Montreal}
  \state{Qu\'ebec}
  \country{Canada}
  \postcode{H3C 1K3}
}

\author{Julien Mercier}
\email{mercier.julien@uqam.ca}
\affiliation{%
  \institution{NeuroLab, D\'epartement d'\'education et formation sp\'ecialis\'ees, \\ Universit\'e du Qu\'ebec \'a Montr\'eal}
  \streetaddress{1 Th{\o}rv{\"a}ld Circle}
  \city{Montreal}
  \state{Qu\'ebec}
  \country{Canada}}

\author{Roberto Erick Lopez-Herrejon}
\email{roberto.lopez@etsmtl.ca}
\affiliation{%
	\institution{\'Ecole de technologie sup\'erieure, Universit\'e du Qu\'ebec}
	\streetaddress{1100 rue Notre-Dame Ouest}
	\city{Montreal}
	\state{Qu\'ebec}
	\country{Canada}
	\postcode{H3C 1K3}
}

\renewcommand{\shortauthors}{Sepasi et al.}

\begin{abstract}
 Software Product Lines (SPLs) are families of related software systems which are distinguished by the set of features each system provides. Feature Models are the de facto standard for modelling the variability of SPLs because they describe the features, their relations, and all the combinations of features that constitute a SPL. Because of their key role, feature models are at the core of many tasks in SPL engineering. Our work presents an empirical study on the comprehension of feature models for the task of checking the validity of configurations. Our study 
 % questions 1 and 2
% \textit{i)} 
 explored the relation between the number of features and the number of cross-tree constraints with the accuracy of participants' answers to validity checking questions, 
 % percentage fixations per questions
 %\textit{ii)} 
 and used eye fixations for analyzing the difficulty in interpreting fixated information and  % percentage fixation time per question
 %\textit{iii)} 
 the amount of cognitive processing of the different parts of the feature model stimuli.
 % parts = AOIs
 % i) results
 We found that answer accuracy does not relate individually to the number of features or to the number of cross-tree constrains of a feature model, but both factors do show an interaction on accuracy. 
 %We also found that incorrect answers took significantly more time. 
 %We found a difference in time between correct and incorrect answers.
 % ii) and iii)
  Additionally, our study identified differences in feature models with cross-tree constraints in both number of fixations and fixation time, but no differences in those models without cross-tree constraints.

%  For the feature models with CTCs we found differences 
 
 % Differences were found both in terms of standardized time spent and standardized number of fixations for problems with CTC, but not for problems without CTC.

 %problem
% 
%  uses eye fixations as metric of difficulty in interpreting fixated information. 
 
% , and explore the relation between correctness of the answers and the cognitive processing of the different aspects of the problem
 
% We also investigated if the correctness of answers could be related to the amount of cognitive processing of the different aspects of the problem. 

 % Note: cognitive load in SE SMS, check if they use proportions. Point out differences in related work. 

 %Our findings indicate that the ... number of features and number of cross-tree constraints ..... The cognitive load traced to the cross-tree constraints on the correctness was  with the correctness performance of the tasks whereas  ... 
\end{abstract}



%
%\begin{CCSXML}
%<ccs2012>
%  <concept_id>10010520.10010553.10010562</concept_id>
%  <concept_desc>Computer systems organization~Embedded systems</concept_desc>
%  <concept_significance>500</concept_significance>
%  <concept_id>10010520.10010575.10010755</concept_id>
%  <concept_desc>Computer systems organization~Redundancy</concept_desc>
%  <concept_significance>300</concept_significance>
%  <concept_id>10010520.10010553.10010554</concept_id>
%  <concept_desc>Computer systems organization~Robotics</concept_desc>
%  <concept_significance>100</concept_significance>
%  <concept_id>10003033.10003083.10003095</concept_id>
%  <concept_desc>Networks~Network reliability</concept_desc>
%  <concept_significance>100</concept_significance>
%</ccs2012>
%\end{CCSXML}
%
%\ccsdesc[500]{Computer systems organization~Embedded systems}


%%% A "teaser" image appears between the author and affiliation
%%% information and the body of the document, and typically spans the
%%% page.

\maketitle

% Sections
% Introduction section
\section{Introduction}
\label{sec:introduction}

Feature models are the de facto standard to represent the features and their combinations whose implementations is used for constructing the set of similar software systems that constitute a \textit{Software Product Line (SPL)}.
Feature models play a crucial role throughout the development cycle of SPL~\cite{SPLE,DBLP:books/FOSP,DBLP:books/sp/variability2013}. Despite their importance, the study of comprehension of these models is an area that remains largely unexplored.
One of the most common tasks performed with feature models is checking that a given configuration, i.e. a combination of features, satisfies or not all the constraints expressed by a feature model. In other words, checking whether a configuration is valid or not. 
Despite the enormous progress in automating the support for this and other similar analysis tasks~\cite{DBLP:journals/csur/ThumAKSS14,DBLP:journals/computing/GalindoBTGR19}, all of them are ultimately driven by SPL developers who must look at a feature model and comprehend its meaning.

The goal of our empirical study is identifying some of the challenges involved in comprehending feature models, more concretely it focuses on the impact that the number of features and the number of cross-tree constraints (see definitions in Section~\ref{subsec:SPL}) have on the comprehension of feature models for answering questions on the validity of configurations. More specifically, our study: \textit{i)} explores the relation of these two feature model metrics with the accuracy of responses and the response time,  and \textit{ii)} uses eye fixations as an estimation of the cognitive effort taken by the different stimuli components of the tasks.   

%
% by examining the effect of important characteristics of the feature models on comprehension. 
%
%This paper presents an empirical ...
%
%Our work contributes an empirical study that focuses on
%

For our experimental design, we calculated the number of features and number of cross-tree constraints for 882 feature models taken from the largest repository available. 
We divided the whole dataset of feature models into four ranges for number of features and 3 ranges for cross-tree constraints, based on the distribution of these metrics along their quartile values.
We obtained a random sample of 2 feature models for each of the twelve factor combinations, yielding a total of 24 feature models for which we anonymized their feature names and devised a validity checking question of similar complexity.   
We recruited 17 participants each of them followed a different random order of tasks. 

Among other findings, our results show that the accuracy of the answers does not relate individually to the number of features or to the number of cross-tree constrains of a feature model. Instead, both of these factors show an interaction in accuracy. More strikingly, we found that looking more frequently or longer at the feature model increases the likelihood of providing an incorrect answer, irrespective of both the number of features and the number of cross-tree constraints. We argue that our work opens several avenues of research on the comprehension of feature models for configuration and other related tasks.

\section{Background}
\label{sec:background}

In this section, we present the basic concepts and terminology of Software Product Lines and Feature Models necessary to follow our empirical study. We introduce a running example that we use to illustrate and explain our study throughout the paper, and we define the core terms related to eye-trackers studies in software engineering. 

\subsection{Software Product Lines and Feature Models}
\label{subsec:SPL}

As mentioned before, Software Product Lines (SPLs) are families of related systems whose members are distinguished by the set of features they provide~\cite{DBLP:journals/tse/BatorySR04,SPLE}. 
There exist an extensive body of research and application of SPL principles that attests the benefits that SPLs provide (e.g.~\cite{SPLE,DBLP:journals/infsof/HeradioPFCH16, DBLP:journals/tse/GalsterWTMA14,DBLP:journals/infsof/ChenB11}).

\begin{figure}[t]
	\centering
	\includegraphics[scale=0.4]{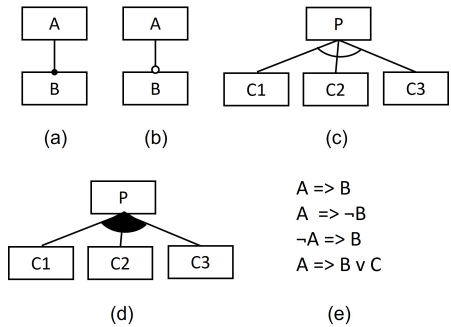}   
	\caption{Feature Models \--- (a) Mandatory feature, (b) Optional Feature, (c) Alternative group, (d) Or group, (e) basic Cross-Tree Constraints}
	\label{fig:fm-basics}   
\end{figure}

In a Software Product Line, each software product is characterized by a different combination of features.
All the possible feature combinations (that correspond to all the software products) are expressed by \textit{variability models} for which there are different alternatives~\cite{DBLP:conf/vamos/CzarneckiGRSW12}; however, feature models have become the de facto standard~\cite{KCH+90}. In this type of model, features are depicted as labelled boxes and their relationships as lines, collectively forming a tree-like structure. The typical graphical notation and a basic textual notation of feature models is shown in Figure~\ref{fig:fm-basics}.

A feature is \textit{selected} when it is present in a software product. A feature 
can be classified as:
\begin{itemize}
\item \textit{Mandatory} which is selected whenever its parent feature is also selected. For instance, in Figure~\ref{fig:fm-basics}(a), if feature \texttt{A}  is selected then feature \texttt{B} must be selected. Mandatory features are denoted with a filled circle on their end of the relationship line with their parent feature.

\item \textit{Optional} which may or may not be part of a product whenever its parent feature is selected. For example, in Figure~\ref{fig:fm-basics}(b), if feature \texttt{A}  is selected, then feature \texttt{B} may or may not be selected. This means that there can be some software products with feature \textbf{B} and some others without it. Optional features are denoted with an empty circle on their end of the relationship line with their parent feature.

\end{itemize}
	
 Features can be grouped into two ways:
 \begin{itemize}
 	\item \textit{Alternative group} whereby if the parent feature of the group is
 	selected, \emph{exactly one} feature from the group must be selected. For example, Figure~\ref{fig:fm-basics}(c), if feature \texttt{P} is selected, then exactly one of the group features \texttt{C1}, \texttt{C2} or \texttt{C3} must be selected. This means that the SPL with this group contains some software products with feature C1 selected, some with feature C2 selected, and some with feature C3 selected. However, no software product of this SPL can have more than one of these 3 features selected as they are mutually exclusive. Alternative groups are denoted with an empty arc across the lines that relate the parent and the children features.  	
 	
 	\item \textit{Or group} whereby if the parent feature of the group is selected, then \emph{at least one} features from the group can be selected. For example, Figure~\ref{fig:fm-basics}(d), if feature \texttt{P} is selected, one or more features among \texttt{C1}, \texttt{C2} or \texttt{C3} must be selected. This means that the SPL with this group can have products with one, two, or three of these features selected. Or groups are denoted with a filled arc across the lines that relate the parent and the children features.  
 	
 \end{itemize}

  In addition to hierarchical parent-child relations explained above, features can also relate across different branches of the feature model with \textit{Cross-Tree Constraints (CTCs)}~\cite{DBLP:journals/is/BenavidesSC10,DBLP:conf/birthday/Benavides19}.
  CTCs can be denoted graphically or using propositional logic formulas because it is a common underlying formal representation for the analysis of feature models ~\cite{DBLP:conf/birthday/Benavides19}. We chose the propositional logic formulas because it is the format used by the FeatureIDE tool\footnote{\url{https://featureide.github.io/}}~\cite{DBLP:books/sp/MeinickeTSBLS17}, that we used to render the visual stimuli for our empirical study as explained in Section~\ref{subsec:stimuli-selection}. 
  
   We employ only a few basic  forms of constraints based on the random sample of feature models selected for our empirical study. More details on the selection of feature models is presented in Section~\ref{subsec:selection-fms}. Our empirical study considers basic CTCs expressed with a logical implication of the form $\alpha => \beta $. We refer to the term $\alpha$ as the antecedent of the implication. Our study considers the following four forms:
  
  \begin{itemize}
  	\item \textbf{\texttt{A => B}}. This form means that if feature \texttt{A} is selected, then feature \texttt{B} must also be selected. 
  	
  	\item \textbf{\texttt{A => \textlnot B}}. This form means that if feature \texttt{A} is selected, then feature \texttt{B} must \emph{not} (symbol \textlnot) be selected.
  	
  	\item \textbf{\texttt{\textlnot A => B}}. This form means that if feature \texttt{A} is \emph{not} selected, then feature \texttt{B} \emph{must} be selected.
  	
  	\item \textbf{\texttt{A => B $\vee$ C}}. This forms means that if feature \texttt{A} is selected, then (as denoted by the $\vee$ operator), either feature \texttt{B}, or feature \texttt{C}, or both must be selected. 
  
\end{itemize}
 
 Finally, because feature models are tree-like structures, they have a \textit{root} feature that is always selected in all the products that conform a SPL. In the next subsection, we provide a running example to illustrate these concepts and our study design in Section~\ref{sec:empirical-study-design}.

 \begin{figure*}[t] 
 	\centering
 	\includegraphics[width=0.8\textwidth]{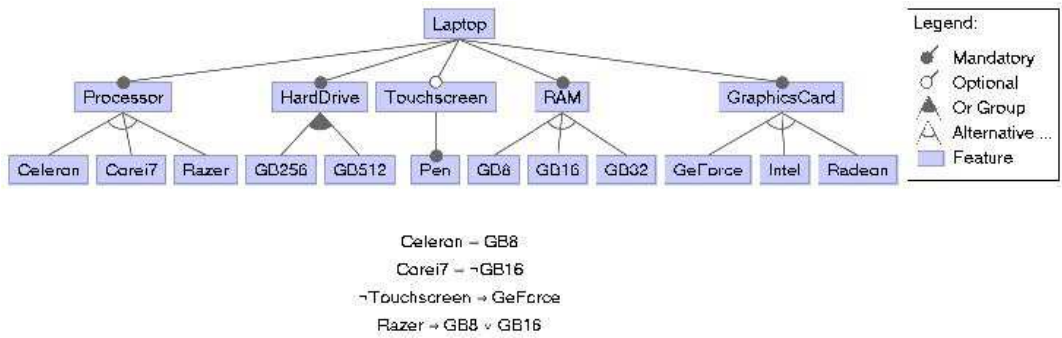}
 	\caption{Running Example \--- Laptop Product Line}
 	\Description{Feature Model with our Running Example}
 	\label{fig:fm-running-example}
 \end{figure*}
 
\subsection{Running Example Description}
\label{subsec:running-example}

We use as a running example a fictitious product line of laptop configurations because it is a domain that is familiar to everybody. This example allows us to illustrate all the notations and concepts used in our empirical study without any loss of generality when applied to the realm of software. 

Figure~\ref{fig:fm-running-example} shows the feature model of our running example. This figure was produced with the FeatureIDE tool. In this product line, a laptop computer is denoted with feature \texttt{Laptop} and has:
\begin{itemize}
	\item A processor, denoted with mandatory feature~\texttt{Processor}. In turn, the laptops can have only one type of processors, represented as an alternative group with the following features: \texttt{Celeron}, \texttt{Corei7}, and \texttt{Razer}.
	
	\item A hard drive, denoted with mandatory feature~\texttt{HardDrive}. In turn, the laptops can have one or two hard drives, which is denoted by an or group with the following features: hard drive of 256GB represented by feature~\texttt{GB256}\footnote{In FeatureIDE, the feature names cannot start with a number, therefore as our naming convention we start with the prefix GB all those features that refer to memory size in GB.}, and hard drive with 512GB represented by feature~\texttt{GB512}.
	
	\item Optionally has a touchscreen, denoted with optional feature~\texttt{Touchscreen}. Hence there are laptops in the product line that have touchscreens and some that have not. For those that have touchscreens, they come with a pen to write with, denoted with mandatory feature \texttt{Pen}. Hence, if a laptop has a touchscreen then it must have a pen.
	
	\item RAM memory, denoted with mandatory feature~\texttt{RAM}. The laptops can have one of three different memory sizes, which are denoted by an alternative group. The options are: 8GB represented by feature~\texttt{GB8}, 16GB represented by feature~\texttt{GB16}, and 32GB represented by feature~\texttt{GB32}.
	
	\item A graphics card, denoted with feature \texttt{GraphicsCard}. The laptops can have only one type of graphics card, which is represented by an alternative group. The options are the following features that correspond to three types of graphics cards: \texttt{GeForce}, \texttt{Intel}, and \texttt{Radeon}.
	
\end{itemize}

In addition to the constraints that derive from the structure of the feature model, our running example also contains the following cross-tree constraints (CTCs): 

  \begin{enumerate}
	\item \texttt{Celeron => GB8}. This CTC means that if a laptop has a \texttt{Celeron} processor then it must have 8GB of RAM, i.e. feature \texttt{GB8} must be selected. 
	
	\item \texttt{Corei7 => \textlnot GB16}. This CTC means that if a laptop has a \texttt{Corei7} processor, then it cannot have a RAM size of 16GB, i.e. feature \texttt{GB16} must \emph{not} be selected.
	
	\item \texttt{\textlnot Touchscreen => GeForce}. This CTC means that if a laptop does not have a touchscreen (i.e. feature \texttt{Touchscreen} is \emph{not} selected), then it must have a GeForce graphics card (i.e. feature \texttt{GeForce} \emph{must} be selected).
	
	\item \texttt{Razer => GB8 $\vee$ GB16}. This CTC means that if a laptop has a \texttt{Razer} processor, then it can either have RAM of 8GB or (operator $\vee$) 16GB, i.e feature \texttt{GB8}, or feature \texttt{GB16}, or both must be selected. 
	
\end{enumerate}

\subsection{Full and Partial Configurations --- Definitions and Examples}
\label{subsec:running-example}

One of the principal tasks that require feature models is configuring a product, that is, defining which features are selected and which are not selected. This configuration task is the focus of our empirical study. Let us now provide working definitions and examples of the terminology we use in our study.

A \textit{configuration} is defined as a set of features selected from a feature model. A configuration can be a\textbf{~\textit{full configuration}} if it considers all the constraints of an entire feature model, or can be a \textbf{\textit{partial configuration}} if it only considers a subset of the constraints of a feature model. Recall that the constraints are those inferred from the hierarchical structure of the feature model (e.g. that only one feature from the alternative group of \texttt{Processor} can be selected) and those written as cross-tree constraints (e.g. if \texttt{Celeron} is selected then \texttt{GB8} must be selected). 

Configurations can be \textit{valid} or \textit{invalid} if they satisfy or not their set of constraints. Thus a \textbf{\textit{valid full configuration}} meets \emph{all} the constraints of an \emph{entire} feature model and conversely an~\textbf{\textit{invalid full configuration}} does \emph{not} meet \emph{all} the constraints of an  \emph{entire} feature model. Similarly for partial configurations. A \textbf{\textit{valid partial configuration}} meets \emph{all} the constraints of a \emph{subset} of the feature model and conversely an~\textbf{\textit{invalid partial configuration}} does \emph{not} meet \emph{all} the constraints of a \emph{subset} of a feature model.
The \textbf{\textit{verification of the validity of a configuration}}, be it partial or full, is defined as the process of checking that all the applicable constraints, hierarchical and cross-tree, are met or not. Let us now provide examples of the verification of the validity of four configurations (two full and two partial).

\begin{sloppypar}
As a first example, please consider the following full configuration with the selected features \texttt{\{Laptop, Processor, Celeron, HardDrive, GB256, Touchscreen, Pen, RAM, GB8, GraphicsCard, Intel\}}. Let us first verify the constraints derived from the structure of the feature model. The root feature \texttt{Laptop} is selected, as well as all its mandatory child features: \texttt{Processor}, \texttt{HardDrive}, \texttt{RAM}, and \texttt{GraphicsCard}. Furthermore, optional feature \texttt{Touchscreen} and its mandatory child feature \texttt{Pen} are both correctly selected. Notice also that only one processor is selected (feature \texttt{Celeron}), one hard drive size is selected (feature \texttt{GB256}), one RAM size is selected (feature \texttt{GB8}), and one graphics card is selected (feature \texttt{Intel}). Now let us focus on the cross-tree constraints. The first CTC is met because feature \texttt{Celeron} is selected and \texttt{GB8} is also selected. The second CTC refers to \texttt{Corei7} in the antecedent of the implication, and because it is not selected (i.e. false) the constraint is met. Similarly for the third CTC, because the \texttt{Touchscreen} feature is selected (i.e. true) the antecedent does not hold and the constraint is met. Similarly also for the fourth CTC whose antecedent is feature \texttt{Razer}, which is not selected and hence the constraint is met. Because this configuration meets all the constraints of the feature model it is a valid full configuration.  
\end{sloppypar}

As a second example, consider the following full configuration with the selected features \texttt{\{Laptop, Processor, Corei7, HardDrive, GB512, Pen, RAM, GB16, GraphicsCard, Intel\}}. Following a similar process as described for the hierarchical constraints of the first example, it is simple to verify that all these constraints are met except one. Namely, feature \texttt{Pen} is selected but its parent feature \texttt{Touchscreen} is not selected. This constraint violation alone is enough to render this configuration as an invalid full configuration. Nonetheless, let us analyze the cross-tree constraints. The first CTC is met because the antecedent of the implication refers to feature \texttt{Celeron} which is not selected. The second CTC is violated because feature \texttt{Corei7} is selected and \texttt{GB16} is also selected in the configuration, situation that is prohibited by this CTC. The third CTC is also violated because given that feature \texttt{Touchscreen} is not selected, this constraint required that feature \texttt{GeForce} were selected, but it is not. The fourth CTC is met because the antecedent refers to feature \texttt{Razer} which is not selected. In summary, this full configuration is invalid because it violates three constraints, one from the hierarchy of the feature model and two cross-tree constraints.   

As a third example, consider the following partial configuration with the selected features \texttt{{Razer, HardDrive, GB256, GB512, Touchscreen, GB8, GeForce}}. Because it is a partial configuration, for its validity we only need to verify the constraints where these features are involved. As before, we verify the hierarchical constraints. Feature \texttt{Razer} is selected and there are no other hierarchical constraints to check for it. Features \texttt{HardDrive}, \texttt{GB256}, and \texttt{GB512} belong to the same or group, hence the constraints of this type of group are met. \texttt{Touchscreen} is a feature that can be selected and there are no other constraints to verify as neither the feature \texttt{Laptop} nor the feature \texttt{Pen} are in the set of features of this partial configuration. Similarly, for features \texttt{GB8} and \texttt{GeForce} there are no other hierarchical constraints to validate as the features they relate to are not selected in the partial configuration. Regarding the CTCs, the first CTC is relevant because it refers to feature \texttt{GB8}. However, because feature \texttt{Celeron} is not in the partial configuration (hence false), the CTC is met.  The second CTC is not relevant as it does not involve any of the features of the partial configuration.
The third CTC is relevant because it refers to two features in the partial configuration. However, because feature \texttt{Touchscreen} is actually selected the antecedent of the implication does not hold and hence the CTC is met. The fourth CTC is met because feature \texttt{Razer} and feature \texttt{GB8} are selected in the partial configuration. In summary, this is an example of a valid partial configuration as it meets all constraints of its subset of the feature model.   
	
As a final example, consider the following partial configuration with the selected features \texttt{\{Corei7, Touchscreen, RAM, GB16, GB32, Intel\}}. Again, because it is a partial configuration we need to verify only the constraints where these features are involved. Feature \texttt{Corei7} is selected and there are no other features that relate to it in the structure of the feature model. Similarly, feature \texttt{Touchscreen} is selected and none of the features it relates to are in the partial configuration. Thus both features meet their subset of hierarchical constraints. Now left focus on the alternative group of feature \texttt{RAM} and the two members of the group, feature \texttt{GB16} and feature \texttt{GB32}, both of which are selected in the partial configuration. This is a clear violation of an alternative group which requires that only one feature in the group can be selected. This violation alone is enough to render make the partial configuration invalid. Nonetheless, let us look at the remaining subset of constraints that need to be verified. Feature \texttt{Intel} is selected but no other feature it relates to are in the partial configuration, thus its constraints are met. Now regarding the CTCs, the first one is not relevant because it does not reference any feature in the partial configuration. The second CTC is violated because \texttt{Corei7} is selected but feature \texttt{GB16} is also selected. The third CTC is met because the antecedent of the implication does not hold as feature \texttt{Touchscreen} is in the partial configuration. The fourth CTC is met because the antecedent of the implication is not in the partial configuration. In summary, this is an invalid partial configuration.  

% 60 configurations are valid.

\subsection{Eye-tracking Basics}
\label{subsect:eye-tracker-terminology}

There is an extensive and long standing body of research in eye-tracking theory and practice (e.g.~\cite{DBLP:books/Duchowski17,DBLP:books/HolmqvistA17}). Among this research, there is a vast number of works in the area of software engineering as summarized, for example, by Obaidellah et al. ~\cite{DBLP:journals/csur/ObaidellahHC18}, and Sharafi et al. ~\cite{DBLP:journals/infsof/SharafiSG15,DBLP:journals/ese/SharafiSGBBC20}. Eye tracking is based on the visual attention of participants whose eye gaze data is recorded. The premise is that visual attention triggers the cognitive processes required for the comprehension and resolution of tasks, and in turn such cognitive processes direct the visual attention to specific locations on the visual field of the participant.  

As defined by Sharafi et al., a \textit{visual stimulus} is any object required to perform a task whose visual perception triggers the participant's cognitive processes to perform some actions related to the task~\cite{DBLP:journals/ese/SharafiSGBBC20}. Because our study focuses on feature models and their comprehension, they are at the core of our visual stimuli. Section~\ref{subsec:stimuli-selection} provides further details on the stimuli definition.
Our empirical study focuses on \textit{fixations} which are eye movements that stabilize the retina on an object of interest in a visual stimulus~\cite{DBLP:books/Duchowski17}. The duration of the fixations varies by the task performed and the participants, with typical values can range from 100 to 400 milliseconds~\cite{DBLP:books/Duchowski17,DBLP:books/HolmqvistA17}. 
%As a visual example, Figure~\ref{fig:example-fixations} shows a small portion of the feature model of our running example with five fixations on different parts of the model. In this figure, fixations are depicted as circles whose radius depend on the duration of the fixation, the larger the radius the longer the duration. In addition, the circles are labelled with a number which indicates the order in which the fixations occurred, constituting a \textit{scan path}.

An \textit{Area Of Interest (AOI)} is an area in the visual stimulus where the researcher is interested in gathering data about because they are deemed relevant for performing the participants' tasks~\cite{DBLP:books/HolmqvistA17}. AOIs are used to defined third order data, according to Sharafi's et al. data classification ~\cite{DBLP:journals/ese/SharafiSGBBC20}. In particular, our study considers two third order data: 1) \textit{fixation count} which is defined as the number of fixations on a specific AOI, 2) and \textit{fixation time} which is defined as the aggregated duration of all the fixations on a specific AOI. In our study, we defined multiple AOIs based on the semantics of the feature models and used ratios of these two measures to normalize the data across participants and questions~\cite{DBLP:books/HolmqvistA17}. Further details on the AOIs and their relation to the research questions are provided throughout the coming section.

\section{Empirical Study Design}
\label{sec:empirical-study-design}

In this section, we state our research questions and provide their rationale, we describe the process to select the feature models of our study, and how we constructed the visual stimuli from them. Furthermore, we describe our experimental design, including the selection of participants and assignment of experimental conditions. 

\subsection{Goal and Research Questions}
\label{subsec:research-questions}

As mentioned before, there is an extensive research on metrics of feature models~\cite{DBLP:journals/infsof/El-SharkawyYS19}. Our empirical study focuses on two of the most basic metrics namely, \textit{Number of Features (NoF)} and \textit{Number of Cross-Tree Constraints (NoC)}. We state our research goal as follows:

\textit{\textbf{Goal}}. To analyze the impact of the metrics \textit{Number of Features (NoF)} and \textit{Number of Cross-Tree Constraints (NoC)} in the comprehension of feature models for the task of verifying the validity of full and partial configurations.

Henceforth, for sake of brevity, whenever we refer to \emph{verification tasks} we imply the verification of the validity of configurations. The research questions of our empirical study are the following: 
%Our empirical study considers \textcolor{red}{four} questions. The first two relate to the accuracy and the response time of the configuration verifications. The last two correspond to the gaze analysis of the cognitive process underneath the comprehension of the feature models while performing the verifications.  

\textbf{\textit{RQ1. Is there any effect of NoF and NoC on the accuracy of verification tasks?}}
\textit{Rationale:} This question examines the effect of NoF, and NoC on the accuracy of the response to a verification task.

%This question aims at establishing if NoF, NoC or both 
%
%... can determine whether the response to a verification task is correct. 
%
%... can influence the accuracy of the response to a verification task.
%
%
%
%... ?given number of correct answer ...

%\item 
\textit{\textbf{RQ2. Is there any effect of NoF and NoC on the response time of verification tasks that were answered correctly?}} 
%Is there any effect of NoF and NoC on the response time of correct instances of configuration validity verifications?
\textit{Rationale:} The purpose of this question is to identify for the tasks answered correctly whether or not there is an effect of NoF or NoC.

% verifying configuration validity 

\textbf{\textit{RQ3. Can the accuracy of verification tasks be inferred from the ratios of fixation counts within specific AOIs of feature models' stimuli?}}
\textit{Rationale:} The ratios of fixation counts within the AOIs of a feature model can be interpreted as a proxy of the cognitive effort taken for the comprehension of such AOI during a verification task. We want to find out if this proxy can be used to infer the accuracy of responses.

%% Previous version
%%\item 
%\textbf{\textit{RQ3. What is percentage of fixations of the eye gaze at different parts of a feature model stimulus while performing validity verifications tasks?}}
%\textit{Rationale:} The percentage of fixations on an element of a feature model can be interpreted as a proxy of the cognitive effort taken for the comprehension of such element during the verification of the validity of a configuration. In this way, the percentages of the different parts of a feature model could help explain both the accuracy and response time of the verifications.
%%\item 

\textit{\textbf{RQ4. Can the accuracy of verification tasks be inferred from the ratios of fixation time within specific AOIs of feature models' stimuli?}}
\textit{Rationale:} Similar to the previous question, we want to find out if the ratio of fixation time within the AOIs of a feature model can be used to infer the accuracy of responses.

\subsection{Selection of Feature Models}
\label{subsec:selection-fms}

For our empirical study, we obtained all the 882 feature models available from SPLOT repository\footnote{\url{http://www.splot-research.org/}. Accessed on July 2021.}, which has the largest dataset of publicly available feature models. With the provided SPLOT API, we implemented the feature metrics NoF and NoC and applied them to our dataset. Table~\ref{tab:fm-distribution}(a) shows the minimum, maximum, and mean values as well as the first, second and third quartile values of our metrics NoF and NoC. On closer inspection, we observed a long-tailed distribution of values of NoC. Based on our experience using feature models, we decided to set the maximum value of NoC at 15 for our study. This threshold eliminated only 35 feature models out of the original 882 models in our dataset. 

We then analyzed the distribution of the feature models along the quartile values for the NoF metric. The ranges considered were: (1) Min<=NoF<Q1, (2) Q1<=NoF<Q2, (3) Q2<=NoF<Q3, and (4) NoF>=Q3. For the NoC metric, we set three ranges: (1)  feature models with no CTCs (i.e. NoC=0, between minimum and Q1), (2) feature models with 1 or 2 CTCs (i.e. Q1<=NoC<Q3), and (3) feature models with 3 to 15 CTC (i.e. Q3<=NoC<=15). 

Table~\ref{tab:fm-distribution}(b) shows the number of feature models for each range combination of NoF and NoC. The highest frequency with value 114 was for feature models with between 20 to 34 features (range 3 in NoF dimension) and without CTC (range 1 in NoC dimension). The lowest frequency with value 29 was for feature models with 10 to 13 features (range 1 in NoF dimension) and 3 to 15 CTC (range 3 in dimension NoC). 
As a final step, for our empirical study we randomly selected two feature models for each combination of ranges along both dimensions
% \footnote{The feature models and the source code used for the random selection will be made available in an open access repository.}
. 
The details of the 24 feature models selected and their NoF and NoC are in our replication package. %The 24 feature models selected and their NoF and NoC are shown in Table~\ref{tab:fm-distribution}(c). 
\begin{table}
	\centering
	\caption{Feature Model Selection Summary}
	\label{tab:fm-distribution}
	\subcaption*{(a) Descriptive statistics, N=882}
	\begin{tabular}{|c||c|c|c|c|c|c|}
	\hline
	Metric	& Min & Q1 & Q2 & Mean & Q3 & Max \\ \hline
	NoF  & 10 & 14 & 20 & 27.95 & 35 & 366 \\ \hline 
	NoC &  0 & 0 & 1 & 3.531 & 3 & 246  \\ \hline
	\end{tabular}
	\bigskip
	\subcaption*{(b) Feature models per range of NoF and NoC, N=847}
	\begin{tabular}{cc|c|c|c|c|}
	\cline{3-6}
	& & \multicolumn{4}{|c|}{\textbf{NoF}} \\ 
	\cline{3-6} 
	%		  \toprule
	%		\hline
	%		  & \multicolumn{4}{|c|}{NOF} \\ \hline 
	& & \textbf{{\small (1) 10..13}} & \textbf{{\small (2) 14..19}} & \textbf{\small {(3) 20..34}} & \textbf{{\small (4) >=35}} \\ 
	\cline{1-6}
	% \midrule
	\multicolumn{1}{|c}{\multirow{3}{*}{\textbf{NoC}}} & \multicolumn{1}{|l||}{\textbf{(1) 0}} & 106 & 93 & 114 & 78 \\
	\cline{2-6} 
	\multicolumn{1}{|c}{} & \multicolumn{1}{|l||}{\textbf{(2) 1..2}} & 73 & 88 & 40 & 34 \\ 
	\cline{2-6}
	\multicolumn{1}{|c}{} & \multicolumn{1}{|l||}{\textbf{(3) 3..15}} & 29 & 27 & 77 & 88 \\ \hline
	%\bottomrule
    \end{tabular}
	Ranges for NoF:1,2,3,4  Ranges for NoC: 1,2,3
% TODO removing the table of feature models examples
%	\bigskip
%	\subcaption*{(c) Sampled Feature Models, N=24}
%	\footnotesize
%%	\label{tab:fm-distribution}
%\begin{tabular}{|c||p{1.5cm}|c|c||p{1.5cm}|c|c|c|}
%	\hline
%	Ranges & Name & NoF & NoC & Name & NoF & NoC \\ \hline  \hline
%	1,1 & {\footnotesize Smart Queue} & 13 &	0 & {\footnotesize poker spl} & 11 & 0 \\ \hline
%	1,2 & Name (1) & 10 &	2 &	Name (2) &	10&	2 \\ \hline
%	1,3 & Xor Example & 13 & 6 & Name (3) & 13 & 6 \\ \hline  \hline
%	2,1 & Stack PL &	17 & 0  & Name (4) & 18 & 0  \\ \hline
%	2,2 & Master & 17	 & 1 & Name (5) & 16 & 1 \\ \hline
%	2,3 & Nexus DSPL & 19 & 7 & FIRE-ALARM &	19&	4 \\ \hline  \hline
%	3,1 & B2B Website & 24 & 0 & ATM Software & 29 & 0  \\ \hline
%	3,2 & MKT	& 22 &	1 & Smart Home &	22 &	2  \\ \hline
%	3,3 & Bike &	21 &	6 &	Name (6) & 32 &	4 \\ \hline  \hline
%	4,1  & Name (7) & 50 & 0 & DS Sample & 41 & 0 \\ \hline
%	4,2 & TestPlant	&65 &	2 &	 Name (8) &	58 &1 \\ \hline
%	4,3 & Inmobiliaria & 39 &	5 &	Software Stack &	38 &	6 \\ \hline
%\end{tabular}
%Range values i,j : i range in NoF, j range in NoC. \\ Names are: (1) SPL\_e\_commerce, (2) Example Mobile Phone, (3) FM with optional XOR, (4) markke core asset testing, (5) MobileMedia-Bruno-ExpUFMG, (6) IVI Development VM, (7) IosAccessibiliteFrench, (8) MayaVariabilityModel  
\end{table}

% NOTE: There were some manual transformation to render them available for FeatureID.

\subsection{Tasks and Stimuli Selection}
\label{subsec:stimuli-selection}

We define a task for each feature model in our sample. A task consist of answering a validity verification question that has one of the following two forms, where \emph{<configuration>} stands for a list of features in a feature model:
\begin{itemize}
	\item Is \emph{<configuration>} a valid partial configuration?\footnote{One  question of this type has a different wording that requires the completion of the configuration.}
	\item Is \emph{<configuration>} a valid full configuration?
\end{itemize}

Please do notice that in feature models, the names of the features are meaningful as they usually convey additional information about the relationships among features. This information could be used while executing tasks with the feature models and it may have a positive or negative effect on the performance depending on the familiarity of the participants with the domain of the feature models. Hence, we anonymized the features' names to prevent any such effect by renaming the features with sequences of letters in alphabetical order and increasing length following a breadth-first traversal order of the feature model.  

For the selection of the questions, we considered a length of six features and favoured full configurations over partial configurations.  In some cases, it was necessary to consider more or less features to provide meaningful questions. 
In summary, 17 questions use a partial configuration and 7 questions a full configuration. Regarding the number of features in the configurations, 14 questions have 6 features, four questions have 7 features, two questions have 3 features, two questions have 8 features, one question has 5 features, and one question has 10 features.

%# 3  5  6  7  8 10 
%# 2  1 14  4  2  1 

\begin{figure*}[t] 
	\centering
	\includegraphics[width=0.8\textwidth]{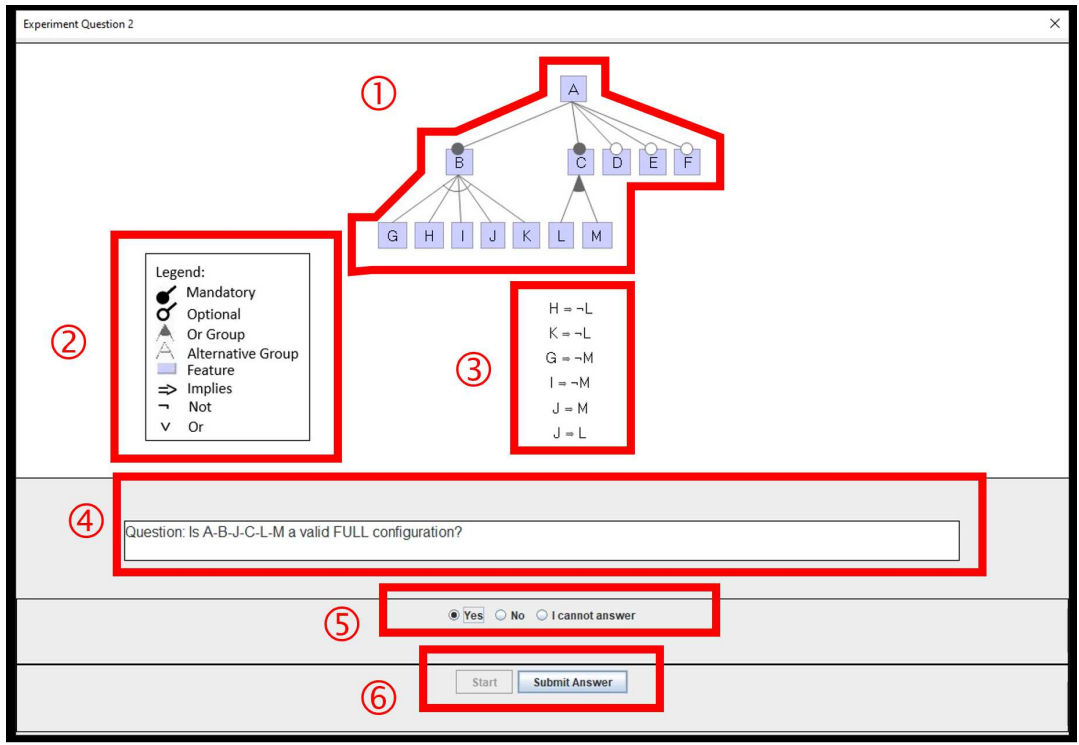}
	\caption{Stimulus Example \--- Java graphical interface with superimposed Areas of Interest (AOIs): \ding{192} feature model (FM), \ding{193} legend label (Label), \ding{194} cross-tree constraints area (CTC), \ding{195} question panel (Question), \ding{196} answer panel (Answer), and \ding{197} buttons panel (Buttons). }
	\Description{Stimulus example of configuration question with AOIs}
	\label{fig:stimulus-example}
\end{figure*}

We also aimed at balancing the distribution of the features that are part of the configuration question across the feature models. 
%the image of the feature models which were rendered using FeatureIDE. 
This means that we selected the features of the configuration questions in different areas along the horizontal axis (i.e., left, middle, right) and the vertical axis (i.e., top, middle, bottom). As an example, 
Figure~\ref{fig:stimulus-example} shows a full configuration question involving 13 features and 6 CTCs.
Notice also that for readability we separated the features in the configuration with a dash (\---) rather than with commas. 
In this figure, three of the features are leaves on the tree-structure, two are children of the root feature, and half of the features appear to the left of the root and half to the right.

%the the anonymized version of the feature model \texttt{DS Sample} with 41 features and no CTCs. Notice also that for readability we separated the features in the configuration with a dash (\---) rather than with commas.  

%Figure~\ref{fig:stimulus-example} shows a partial configuration question involving 6 features for the the anonymized version of the feature model \texttt{DS Sample} with 41 features and no CTCs. 

We explored different alternatives to present the stimuli of our study, among them iTrace\footnote{\url{https://www.i-trace.org/}} and PsychoPy\footnote{\url{ https://www.psychopy.org/}}. However, it was not possible to link FeatureIDE, which is an Eclipse plugin, with them to allow us to obtain the information required for our study. Therefore, we wrote a bespoken Java program that presents a configurable sequence of questions, like that of Figure~\ref{fig:stimulus-example}.
 
The program provides a simple user interface with the following elements:
 \textit{i)} the feature model, \textit{ii)} a legend that summarizes the notations\footnote{We manually extended the legend provided by default by FeatureIDE to include the notation of CTCs and placed it at the same position in all the questions.}, \textit{iii)} area for cross-tree constraints of the form presented in Section~\ref{subsec:SPL}, 
 \textit{iv)} a question panel where the question is presented, \textit{v)} an answer panel that shows the three possible answers \texttt{Yes}, \texttt{No}, and \texttt{I cannot answer} (given to participants to indicate that they were not able to respond), and \textit{vi)} a response panel with the buttons to start answering the question and submitting the answer. 
 The program records the responses and the response time.
 The feature model appears once the participant hits the 
 \texttt{Start} button and the response time clock gets started. 
 %At the beginning of each question, only the \texttt{Start} button is enabled and the feature model is not shown.
%  When the \texttt{Start} button is hit, the feature model appears and the response time clock gets started. 
  The response time ends when the participant hits the \texttt{Submit Answer} button which also clears the feature model for the next question. 
 
 %The source code of the Java program as well as the 24 feature models, their questions and images are all available in our repository.
 
% TODO Note: Better use one example with CTCs

%of the following forms:a question of vali The tasks hav
%
%Anonymization of features. Addition of legend.
%Selection of the questions.
%
%NOTE: These questions will be part of the refinements of the four RQ after the design of the stimuli is presented.
%
%Explain the results observed in RQ1 to RQ3 using eye-tracking input
%
%Prevalence .. how much of something…
%
%RQ4. What is percent of time spent on consulting legend of the feature model?
%
%RQ5. What is percent of time spent on consulting the answer panel?
%
%RQ6. What is percent of time spent on consulting the CTC area?
%
%Cognitive process.
%
%Frequency of fixations= number of fixations per AOI / all fixations 
%
%RQ7. What is the frequency of fixations in the legend of the feature model?
%
%RQ8. What is the frequency of fixations in the answer panel?
%
%RQ9. What is the frequency of fixations in the CTC area?

\subsection{Experimental Design}
\label{subsec:experimental-design}

We followed the guidelines proposed by Sharafi et al. for planning and executing our empirical study~\cite{DBLP:journals/ese/SharafiSGBBC20}. We chose a within-subjects design where each participant performed the 24 tasks, each task being a validity verification question as described above. In addition, we randomized the order of the tasks in such a way that each participant follow a different order~\cite{DBLP:books/LFH2017}.

We recruited 19 participants, two of them piloted our experimental design, mostly graduate students from our institution and without prior knowledge of feature models. We created a tutorial video to fill this background gap. The video is an extended and detailed version of the contents presented from Section~\ref{subsec:SPL} to Section~\ref{subsec:running-example}. This video has a duration of about 40 minutes and was presented to each participant right before the start of his/her tasks.
%\footnote{The video cannot be annoymized but will be made available as part of our replication package.}.

We performed our study in a research lab  devoted to user studies at our institution, where a participant is alone in the experiment room and the experimenter is at an adjacent observation room overseeing the experiment. This arrangement guaranteed a stable and regular environment without any distractions, with controlled factors such as lighting of the room and position of the participants in relation to the screen, keyboard and eye-tracker. For our study, we used the Tobii Pro Fusion bar eye-tracker and Tobii Pro Lab tool for capturing and analyzing the data. The session of each participant followed the sequence:
\begin{itemize}
	 \item Brief introduction of the experiment by the experimenter, describing its goal and the protocol to follow.
	 \item Signing the consent form in accordance with the ethics certificate of our institution.
	 \item Watching training video and clarifying any questions or issues that the participants may have.
	 \item Calibration of the eye-tracker using Tobii Pro Lab tool.
	 \item Warm-up practice with two questions in the Java application (see Section~\ref{subsec:stimuli-selection}) to gain familiarity with the graphical interface.
	 \item Semi-structured interview to gather further insights from the participants. The interviews were recorded for subsequent analysis.
\end{itemize}	

More formally, our empirical study is a factorial design with two factors that correspond to our independent variables~\cite{DBLP:books/ExperimentationSoftEngWolinRHOR12}. The first factor is Number of Features (NoF) with 4 categorical values, and the second factor is Number of Cross-Tree Constraints (NoC) with 3 categorical values, respectively corresponding to their number of ranges as explained in Section~\ref{subsec:selection-fms}. 

Recall that our Java interface collects two pieces of information for each question: i) correctness, with values \texttt{true} and \texttt{false} to indicate whether the response was correct or not, and b) response time in milliseconds, i.e. the elapsed time between clicking on the \texttt{Start} button and hitting the \texttt{Submit Answer} button. Thus, these are our first two dependent variables, respectively in nominal and ratio scales. They are used for answering our first two research questions, RQ1 and RQ2.

With the aid of Tobii Pro Lab, we defined several areas of interest (AOIs) on top of the Java interface, see Figure~\ref{fig:stimulus-example}. The AOIs roughly correspond to the elements in the graphical interface:~\ding{192} feature model (FM),~\ding{193} legend label (Legend),~\ding{194}cross-tree constraints area (CTC),~\ding{195} question panel (Question),~\ding{196} answer panel (Answer), and~\ding{197} buttons panel (Buttons). 
We should note that the AOI that corresponds to the feature model, i.e. ~\ding{192} in the figure, is tailored to each of the 24 feature models we used in our study. Similarly, the area of interest of CTCs, \ding{194} in the figure, was removed when there were no CTCs and adapted according to the number of CTCs of the feature model.  
The rest of the graphical elements were fixed on the same area of the interface across all questions. For all our AOIs, we collected the fixations count and the fixation time on them. We normalized these values as percentages to analyze across questions and participants~\cite{DBLP:books/HolmqvistA17}. We used these data for the research questions RQ3 and RQ4.
\section{Results and Analysis}
\label{sec:results}

In this section, we start with a general overview of the results obtained before delving into the detailed analysis for each of our four research questions.

\subsection{Descriptive Statistics}
\label{subsec:descriptive-statistics}

\begin{figure}[t] 
	\centering
	\begin{subfigure}[b]{0.45\textwidth} % 0.3\textwidth
		\centering
		\includegraphics[width=\textwidth]{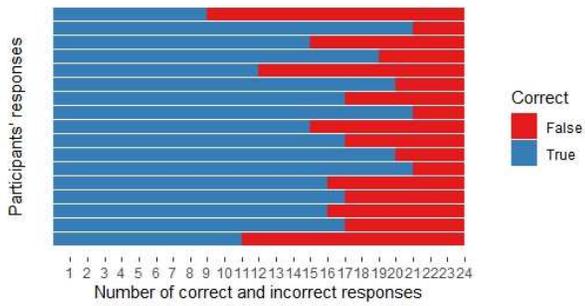}
		\caption{Correct/Incorrect responses per participant}
		\label{fig:responses-participant}
	\end{subfigure}
	\hfill
	\begin{subfigure}[b]{0.5\textwidth}
		\centering
		\includegraphics[width=\textwidth]{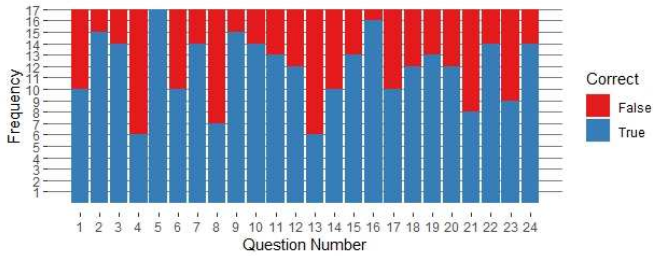}
		\caption{Correct/Incorrect responses per question}
		\label{fig:responses-question}
	\end{subfigure}
%	\hfill
%	\begin{subfigure}[b]{0.45\textwidth}
%		\centering
%		\includegraphics[width=\textwidth]{./figs/correct-answers-boxplot-response-time}
%		\caption{Response time distribution of correct answers}
%		\label{fig:distibution-correct-time}
%	\end{subfigure}
%	\hfill
%	\begin{subfigure}[b]{0.45\textwidth}
%		\centering
%		\includegraphics[width=\textwidth]{./figs/incorrect-answers-boxplot-response-time}
%		\caption{Response time distribution of incorrect answers}
%		\label{fig:distibution-incorrect-time}
%	\end{subfigure}
	\caption{Summary of correcct and incorrect responses}
	\Description{Descriptive Statistics}
	\label{fig:correct-incorrect-responses}
	
\end{figure}

Our empirical study had 17 participants, 10 male and 7 female.
All participants completed the 24 questions, and only one participant in one question indicated that he/she could not answer the question. Figure~\ref{fig:correct-incorrect-responses} summarizes the correct and incorrect responses per participant and per question.

Figure~\ref{fig:responses-participant} shows for each participant the number of correct and incorrect responses. In total for all participants, there were 284 correct and 124 incorrect answers. 
Participants had between 9 to 21 correct answers, with a median of 17, and conversely they had between 3 and 15 incorrect answers with a median of 7.
On average, the correct answers took 42.84 seconds to respond with a standard deviation of 26.15 seconds. For the incorrect answers, the average response time was 52.96 seconds, with a standard deviation of 37.82 seconds. 
Figure~\ref{fig:responses-question} shows the distribution of correct and incorrect answers per question. The minimum was six correct answers for questions 4 and 13, and the maximum was 17 for question 5 which all the participants answered correctly. The average was 11.83 correct answers per question with a standard deviation of 3.08.

\begin{figure}[]  %t 
	\centering
%	\begin{subfigure}[b]{0.45\textwidth} % 0.3\textwidth
%		\centering
%		\includegraphics[width=\textwidth]{./figs/barchart-participants-correct-incorrect-responses}
%		\caption{Correct/Incorrect responses per participant}
%		\label{fig:responses-participant}
%	\end{subfigure}
%	\hfill
%	\begin{subfigure}[b]{0.5\textwidth}
%		\centering
%		\includegraphics[width=\textwidth]{./figs/qn-correct-incorrect-frequency}
%		\caption{Correct/Incorrect responses per question}
%		\label{fig:responses-question}
%	\end{subfigure}
%	\hfill
	\begin{subfigure}[b]{0.45\textwidth}
		\centering
		\includegraphics[width=\textwidth]{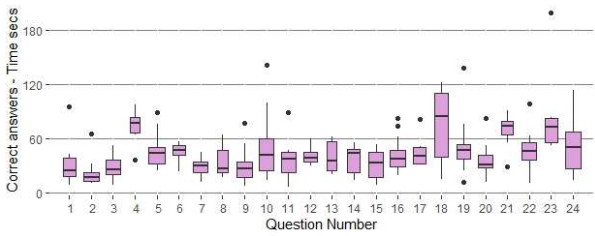}
		\caption{Response time distribution of correct answers}
		\label{fig:distibution-correct-time}
	\end{subfigure}
	\hfill
\begin{subfigure}[b]{0.45\textwidth}
	\centering
	\includegraphics[width=\textwidth]{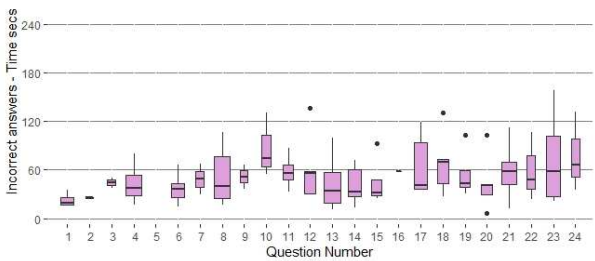}
	\caption{Response time distribution of incorrect answers}
	\label{fig:distibution-incorrect-time}
\end{subfigure}
	\caption{Response time distribution per question}
	\Description{Descriptive Statistics}
	\label{fig:response-time-distribution}
	
\end{figure}

Figure~\ref{fig:response-time-distribution} summarizes the distribution of response time per question for correct and incorrect answers.
Figure~\ref{fig:distibution-correct-time} shows the distribution of response time of correct answers per question. Notice that only four questions have their third quartile above one minute. One question above this threshold is question 4 which was one with the least number of correct answers.
In contrast, Figure~\ref{fig:distibution-incorrect-time} shows the distribution of response time of incorrect answers per question. For the majority of the questions, the distribution is more widely spread, with only 3 questions whose their third quartile was over the one minute threshold.

% Above: 4 (Full, 6). 18 (Partial, 7 ) . 21 (Partial,6) . 23 (Partial, 8) . 24 (6, Partial)

% Summary of correct answers per participant
%  Min. 1st Qu.  Median    Mean 3rd Qu.    Max. 
%9.00   15.00   17.00   16.71   20.00   21.00 

% Summary of incorrect answers per participant
% Min. 1st Qu.  Median    Mean 3rd Qu.    Max. 
% 3.000   4.000   7.000   7.294   9.000  15.000

% Summary of correct answers per question
%   Min. 1st Qu.  Median    Mean 3rd Qu.    Max. 
% 6.00   10.00   12.50   11.83   14.00   17.00 

\begin{figure}[t] 
	
	%\centering
	%\includegraphics[width=0.8\textwidth]{./figs/stimuli-example-annotated}
	%	\includegraphics[width=\linewidth]{./figs/laptop-pl-complete-legend}
	% \caption{Descriptive Statistics}
	
	% scale=0.7
	
	\centering
	\begin{subfigure}[b]{0.5\textwidth}
		\centering
		\includegraphics[width=\textwidth]{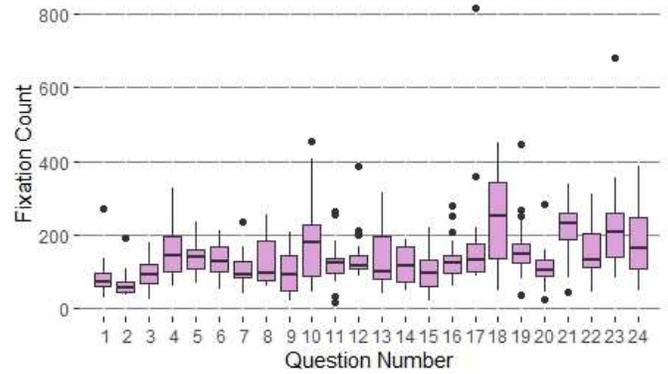}
		\caption{Fixation count per question}
		\label{fig:fixations-per-question}
	\end{subfigure}
	\hfill
	\begin{subfigure}[b]{0.5\textwidth}
		\centering
		\includegraphics[width=\textwidth]{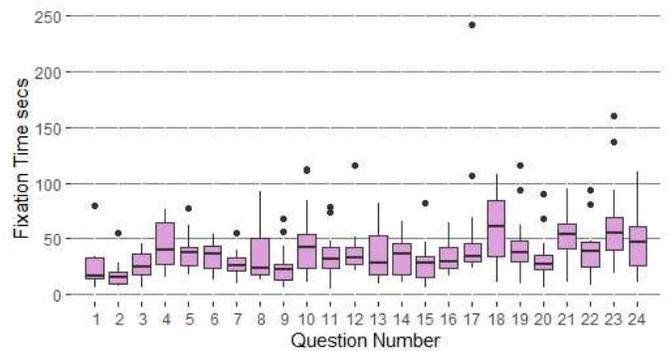}
		\caption{Total fixation time per question}
		\label{fig:total-fixation-time}
	\end{subfigure}
	%	\hfill
	%	\begin{subfigure}[b]{0.45\textwidth}
	%	\centering
	%	\includegraphics[width=\textwidth]{./figs/histogram-aois-fixations}
	%	\caption{Percentage of fixations per question grouped by AOI}
	%	\label{fig:hist-fixations}
	%	\end{subfigure}
	%	\hfill
	%	\begin{subfigure}[b]{0.45\textwidth}
	%	\centering
	%	\includegraphics[width=\textwidth]{./figs/histogram-aois-fixation-time}
	%	\caption{Percentage of fixation time per question grouped by AOI}
	%	\label{fig:hist-fixation-time}
	%	\end{subfigure}
	\caption{Fixations count and fixation time per question}
	\Description{Fixations count and fixation time per question}
	\label{fig:fixations-and-time}	
\end{figure}

Figure~\ref{fig:fixations-and-time} summarizes our findings in terms of fixation count and fixation time per question. Across all participants, the quartile values for fixation counts were: 84, 124, and 184. The minimum value was 18, the maximum value was 817 and the standard deviation was 92.31. Along the same lines, the quartile values of fixation time per question were: 21.11 sec, 32.5 sec, and 48.05 sec. The minimum value was 5.14 sec, the maximum 242.29, and the standard deviation of 25.85 sec. 
Figure~\ref{fig:fixations-per-question} shows the distributions of fixation count per question. Except for questions 18, 21, 23, and 24, all the other questions have median values below 200 fixations. Notice as well the presence of outlier values, in particular in question 17 and question 23. Figure~\ref{fig:total-fixation-time} shows the aggregated fixation time per question. This figure follows a similar pattern with the same questions having medians above 50 seconds of total fixation time. 

\begin{figure}[t] 	
	\centering
%	\begin{subfigure}[b]{0.5\textwidth}
%		\centering
%		\includegraphics[width=\textwidth]{./figs/number-fixations}
%		\caption{Number of fixations per question}
%		\label{fig:fixations-per-question}
%	\end{subfigure}
%	\hfill
%	\begin{subfigure}[b]{0.45\textwidth}
%		\centering
%		\includegraphics[width=\textwidth]{./figs/fixation-time}
%		\caption{Total fixation time per question}
%		\label{fig:total-fixation-time}
%	\end{subfigure}
%	\hfill
	\begin{subfigure}[b]{0.45\textwidth}
	\centering
	\includegraphics[width=\textwidth]{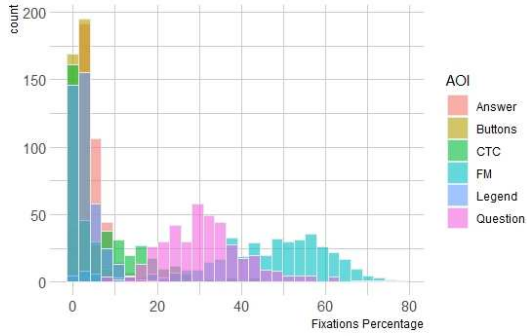}
	\caption{Ratios of fixations count per question grouped by AOI}
	\label{fig:hist-fixations}
	\end{subfigure}
	\hfill
	\begin{subfigure}[b]{0.45\textwidth}
	\centering
	\includegraphics[width=\textwidth]{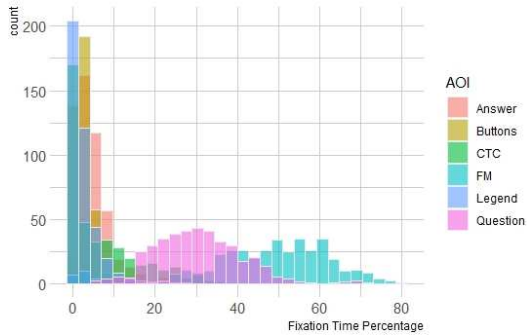}
	\caption{Ratios of fixation time per question grouped by AOI}
	\label{fig:hist-fixation-time}
	\end{subfigure}
	\caption{AOI fixation count and fixation time results}
	\Description{AOI fixation count and fixation time results}
	\label{fig:aoi-fixations-and-time}	
\end{figure}

Recall that in Section~\ref{subsec:stimuli-selection},  we identified six AOIs in the stimuli corresponding to each question of our study. Figure~\ref{fig:aoi-fixations-and-time} shows the results of fixation count and fixation time by AOI. 
Figure~\ref{fig:hist-fixations} shows a stratified histogram of the ratio on the number of fixations per question grouped by AOI. The values of the mean and standard deviation for each AOI are as follows: Feature Model ($\overline{x}=45.27$, $s=15.97$), Question ($\overline{x}= 31.66$, $s=10.50$), CTC ($\overline{x}=7.273$, $s=8.76$), Answer ($\overline{x}=4.37$, $s=3.08$), Legend ($\overline{x}=3.11$, $s=3.51$), and Buttons ($\overline{x}=2.05$, $s=1.93$). All these values are reflected in the figure, whereby the histogram values of the feature model (FM) appear at the rightmost part of the graph, followed to the left by the histogram values of the Question and CTC AOIs respectively. The ratio  contribution of AOIs Buttons, Legend and Answer were rather small as depicted with the highest frequency on the left of the graph.
%# Fixations
%# FM 0.4527 0.1597964
%# Question  0.316610.1050689
%# CTC 0.07273 0.08761329
%# Answer 0.04374 0.03086291
%# Legend 0.031134 0.03517941
%# Buttons 0.020553  0.01936261 
A very similar scenario was observed on the ratio of fixation time per question grouped by AOI shown in Figure~\ref{fig:hist-fixation-time}. The values for this figure are: Feature Model ($\overline{x}=46.19$, $s=16.87$), 
Question ($\overline{x}=30.66$, $s=10.93$), 
CTC ($\overline{x}=7.252$, $s=9.18$), 
Answer ($\overline{x}=5.693$, $s=043.70$), 
Legend ($\overline{x}=2.66$, $s=3.64$), and Buttons ($\overline{x}=2.67$, $s=2.64$).
Here again, the histogram values of the feature model (FM) have the highest ratios values to the right of the graph, followed by Question and CTC. Once again, the highest counts were for AOIs Legend, Answer and Buttons.

\subsection{RQ1 results}
\label{subsec:rq1-results}
Recall from Section~\ref{subsec:research-questions} that this question aims at identifying any effects of NoF and NoC on task accuracy. Hence, the independent variables are the NoF and NoC that are of categorical scale, and the dependent variable is task accuracy (with correct and incorrect values). 
To answer this question, we used multiway frequency analysis whereby the frequency of a cell (i.e., a combination of factors) is the dependent variable that is influenced by one or more categorical factors and their associations ~\cite{multivariate-stats}.
In this method, the null hypothesis $H_{0}$ is that the factors do not affect the frequency distribution of a dependent variable. 

%In our case the experimental factors are NoF and NoC and the dependent variable is accuracy (rate?).
%For us, The test involved success rate as the dependent variable, and NoF and NoC as experimental factors.
 
Our results show that there are neither main effects of NoF nor of NoC on task accuracy. However, there is an interaction effect of NoF and NoC on task accuracy ($\chi^{2}_6 = 21.82$, $p <.001$).
 The higher the value of NoC caused a lower task accuracy, but this drop is different across the values of NoF.
 This pattern of results can be interpreted as an indication that the value of NoC 
imposes more cognitive challenges than the value of NoF.

\subsection{RQ2 results}
\label{subsec:rq2-results}

Recall from Section~\ref{subsec:research-questions} that for this question, we are interested in the effect of NoF and NoC on the response time of tasks answered correctly. An ANOVA test was used to establish the effect of two categorical factors, NoF and NoC, on a continuous dependent variable, response time. The null hypothesis $H_{0}$ is that the means across both factors and their interaction are equal~\cite{multivariate-stats}.  
% Note: if we make distinction between correct and incorrect
%Firstly, we found a difference in response time between correct and incorrect answers ($F_{1,407} = 9,72$, partial $\eta^2= 0.023, p < 0,002$). Incorrect answers are associated with longer response times.
For correct answers, both NoF ($F_{3,272} = 8.28$, $p < 0.001$, $\eta^2_{p}= 0.084$) and NoC ($F_{2,272} = 11.58$, $p <0.001$, $\eta^2_{p}=0.078$) have a main effect on response time. There is not interaction effect of NoF and NoC on response time. NoF causes an almost steady increase in time needed to answer correctly, and the same is true for NoC. 

% TODO include if the focus ...
%??? For incorrect answers (i.e. value 1), both NoF ($F_{3,112} = 3.47$, partial $\eta^2= 0.085, p < 0.019$) 
%and NoC ($F_{2,112} = 3.65$, partial $\eta^2 = 0.061, p <  0,03$) have a main effect on time on task.

%TODO Note: another question, is there any difference in response time between incorrect vs correct answers.

%For incorrect answers (i.e. value 1), both NoF (𝐹3,112 =3.47, partial 𝜂2 =0.085,𝑝 < 0.019) and NoC (𝐹2,112 =3.65, partial 𝜂2 =0.061,𝑝 <0,03) have a main effect on time on task.

%% Original - previous version
%%There is a difference in time on task between correct/incorrect answers ($F_{1,407} = 9,72$, partial $\eta^2= 0.023, p < 0,002$). 
%For incorrect answers (i.e. value 1), both NoF ($F_{3,112} = 3.47$, partial $\eta^2= 0.085, p < 0.019$) 
%and NoC ($F_{2,112} = 3.65$, partial $\eta^2 = 0.061, p <  0,03$) have a main effect on time on task. 
%For correct answers (i.e. value 2), both NoF ($F_{3,272} = 8.28$, partial $\eta^2= 0.084, p < 0.001$) and NoC ($F_{2,272} = 11.58$, partial $\eta^2=0.078, p <  0,001$) have a main effect on time on task. There is not interaction effect of NoF and NoC on time on task. 

\subsection{RQ3 results}
\label{subsec:rq3-results}

Recall from Section~\ref{subsec:research-questions}, that for this question we want to find out if the ratio of fixation counts of the AOIs can be used to explain the accuracy of the responses. Consequently, the analysis focuses primarily on the accuracy of answers, with the eventual examination of interaction effects of this factor with NoF and NoC. Thus, accuracy, NoF and NoC are the independent variables and the ratios of fixation counts of the six AOIs are our dependent variables, that is, the AOIs Answer, Buttons, CTC, FM, Legend and Question as explained in Section~\ref{subsec:experimental-design}. Because tasks with CTCs involve the additional AOI CTC, we applied the MANOVA test twice, once for the tasks with CTCs and once for the tasks without CTCs~\cite{multivariate-stats}.

For tasks with CTCs, the MANOVA test considered the factors (2:accuracy x 4:NoF x 3:NoC). For this test, the null hypothesis $H_{0}$ for the omnibus test states that the factors are not related to the variance of the dependent variables taken globally. When the omnibus test is significant for one factor or a combination of factors, the null hypothesis $H_{0}$ for each dependent variable is that this factor or combination of factors is not related to the variance of any of the six dependent variables. 
We found from the omnibus test that the accuracy of answers was significant globally, that is, the ratio of fixation counts of each AOI is related to the accuracy of answers ($F_{6,251} = 3.458$, $p<.003$, $\eta^2_{p} = .076$), with no interaction effect of this factor with NoF and NoC.
Because of this finding, we analyzed the implication of accuracy on each of the six dependent variables. We observed significant differences in the means of the following AOIs: FM ($F_{1,256} = 19.150$, $p<.001$, $\eta^2_{p} = .070$), CTC ($F_{1,256} = 8.088$, $p<.005$, $\eta^2_{p}  = .011$), Question ($F_{1,256} = 17.723$, $p< .001$, $\eta^2_{p} = .065$), and Buttons ($F_{1,256} = 4.032$, $p<.046$, $\eta^2_{p} = .016$). 
Differences in the means of fixation count ratios for given AOIs for correct vs incorrect responses were as follows: FM 39\% vs 48\%, CTC 12\% vs 9\%, Question 35\% vs 28\%, and Buttons 2\% vs 1.7\%. This finding means that devoting more cognitive effort in comprehending the FM AOI is related to a higher likelihood of providing an incorrect answer. Inversely, devoting more cognitive effort in understanding the AOIs CTC, Question, or Buttons is related to a higher likelihood of providing a correct answer.

For tasks without CTCs, the MANOVA test considered the factors (2: accuracy x4: NoF) because the value of CTC is now a constant. The results show that the null hypothesis of the omnibus test could not be rejected ($F_{5,124} = 0.296$, $p>.05$), therefore the results for each of the dependent variables were not examined. In other words, the proportion of fixation counts of each AOI is not related to the accuracy of answers. 

%rejected ( F5,124 = 0.296, p > .05), therefore 
% scratch ($F_{3,112} = 3.47$, partial $\eta^2= 0.085, p < 0.019$) 

\subsection{RQ4 results}
\label{subsec:rq4-results}

This question is similar to the previous one, but here we want to find out if the ratio of fixation time of any AOI can be used to explain the accuracy of the responses. Thus, we answered the question in the same way but using the ratios of fixation time.  Again, the tests were run separately for tasks with and without CTC because of the same reasons.

For tasks with CTC, the proportion of time fixating each AOI is related to the accuracy of answers ($F_{6,251} = 3.027$, $p<.007$, $\eta^2_{p} = .067$). Additionally, there is no interaction effect of accuracy of answers with the experimental factors, NoF and NoC. Regarding the relation between accuracy of answers and each of the specific dependent variables (AOIs), the tests are significant for: FM ($F_{1,256} = 15.952$, $p<.001$, $\eta^2_{p} = .059$), Question ($F_{1,256} = 16.486$, $p<.001$, $\eta^2_{p} = .061$), and Answer ($F_{1,256} = 4.796$, $p<.029$, $\eta^2_{p} = .018$). 
Differences in the means of fixation time ratios for given AOIs for correct vs incorrect responses were as follows: FM 40\% vs 49\%, Question 34\% vs 28\%, and Answer 6\% vs 4\%.
This finding means that devoting more cognitive effort in comprehending the FM AOI is related to a higher likelihood of providing an incorrect answer. Inversely, devoting more cognitive effort in understanding the AOIs  Question or Answer is related to a higher likelihood of providing a correct answer.

%Correct answers are related with fixating the FM relatively less (40\% vs 49\%), Question relatively more (34\% vs 28\%), and Answer relatively more (6\% vs 4\%). 

For tasks with no CTC, the proportion of time fixating each AOI is not related to the accuracy of answers ($F_{5,124} = 0.197$, $p > .05$). Since the null hypothesis for the omnibus test could not be rejected, no further statistical analysis was required. 
%no other statistics were consulted.

\subsection{Threats to Validity}
\label{subsec:threats-validity}

Our empirical study faced several potential threats to validity as similar studies that rely on eye-tracker measures~\cite{DBLP:books/ExperimentationSoftEngWolinRHOR12,DBLP:journals/ese/SharafiSGBBC20}. Concerning \emph{internal validity}, a threat was \emph{participant selection}. We addressed this issue by recruiting as large and diverse a set of participants as possible, both in terms of gender and technical knowledge of the subject. Because we used a within-subjects design we faced an \emph{order effect bias} whereby participants perform better as they performed more tasks. We mitigated this threat by using a random order of tasks for each participant. The \emph{instrumentation} threat concerns the quality and reliability of the artifacts used in the experiment. We tackled this threat by following a rigorous protocol that included multiple checks from eye-tracker measures and calibrations for each participant to extensive consistency manual and automated validations of the collected and synthesized data.

Regarding \emph{external validity}, a fundamental threat is the selection of the feature models. Certainly, selecting other data set of feature models might yield different results from ours. However, we argue that we addressed this threat by choosing feature models from the largest repository available and in a random way across the twelve combinations of our two factors that are our feature metrics. Because the majority of our participants were graduate students without background in feature models, we cannot generalize our findings for other groups like experienced developers or variability experts. Nonetheless, this aspect is part of our future work. In terms of \emph{construct validity}, we used standard measures of cognitive load such as accuracy, response time, ratios of fixation counts and fixation time on AOIs which were analyzed with standard statistics procedures~\cite{DBLP:books/HolmqvistA17,DBLP:journals/infsof/GoncalesFS21}. Considering other eye-tracking and multimodal measures of cognitive load are also part of our future work.  

\section{Related Work}
\label{sec:related-work}

There is an extensive body of research in program comprehension analyzed with eye-trackers~\cite{DBLP:journals/csur/ObaidellahHC18,DBLP:journals/infsof/SharafiSG15,DBLP:journals/ese/SharafiSGBBC20}. In this section, we briefly summarize the pieces of work closest to ours.
There are several studies that use eye-trackers on models  of software. For instance, Jeanmart et al. study the impact of the presence or absence of the explicit Visitor pattern in a UML class diagram for comprehension and maintenance tasks~\cite{jeanmart_impact_2009}. They found that the Visitor pattern did not reduce the efforts for the comprehension task but instead that the familiarity with UML and patterns had an impact on both tasks. Similarly, other aspects of UML diagrams and their impact on comprehension have also been studied, e.g. color and layout~\cite{yusuf_assessing_2007}. 

%The analysis of the data collected using an eye-
%tracker showed that the Visitor pattern does not reduce

There is an extensive and seminal line of work by Janet Siegmund and colleagues in the study of program comprehension in code with variability, studying aspects such as color, conditional compilation versus feature-oriented programming, or disciplined vs non-disciplined annotations (e.g,~\cite{DBLP:journals/ese/FeigenspanKALSDPLS13,DBLP:journals/ese/SantosMASA19,DBLP:conf/gpce/SchulzeLSA13}). However, to the best of our knowledge, for this line of work they do not employ eye-tracker measures. In contrast, Melo et al. performed an experiment to understand how developers debug programs with and without variability~\cite{DBLP:conf/iwpc/MeloNHBW17}. In code fragments with variability, they found increases in debugging time and in the number of gaze transitions between the use and definitions of fields and methods.  
Da Costa et al. studied C code with annotations and the impact of three refactorings in terms of program comprehension and visual effort~\cite{daCosta2021}. They found that two of such refactorings had an impact. In contrast with these works on variability, our study uses feature models, not source code, and our task is comprehension of configuration validity rather than code debugging or refactoring. 

\section{Conclusions and Future Work}
\label{sec:conclusions}

We presented the first empirical study on feature model comprehension using eye-tracking measures for tasks of verifying validity in partial and full configurations. Our focus was on the impact of two feature model metrics, number of features (NoF) and number of cross-tree constraints (NoC), on these verification tasks. We found that these tasks are complex from a cognitive point of view because they are affected by multiple factors.
For instance, tasks with more NoF or more NoC involved more cognitive operations to answer correctly as shown by a steady increase on response time.
Also, when considering the different aspects (AOIs) of the visual stimuli, our most striking finding was that looking more frequently or longer at the feature model (AOI FM) of a task increases the likelihood of providing an incorrect answer, irrespective of the NoC and NoF. 
%TODO should it be included?
%We advanced some hypothesis to explain this phenomena that deserve further detailed investigation.

We believe our work opens several avenues for further research. Among them are: studying other eye-tracking measures of cognitive load from  pupil size to EEG~\cite{DBLP:journals/infsof/GoncalesFS21}, employing finer grain AOIs for the feature model AOI that would permit us to identify transition patterns or navigation strategies among AOIs in the form of scan paths and conditional probabilities, analyzing the impact of other feature model metrics (e.g., branching factor), and including a larger and more diverse participant population.

\begin{acks}
% Here we will acknowledge the funding agencies of our work.
This work is partially supported by the Natural Sciences and Engineering Research Council of Canada (NSERC) grant RGPIN-2017-05421.
\end{acks}


\bibliographystyle{ACM-Reference-Format}
\bibliography{references}

\end{document}